# TPU v4: An Optically Reconfigurable Supercomputer for Machine Learning with Hardware Support for Embeddings

Industrial Product*

Norman P. Jouppi, George Kurian, Sheng Li, Peter Ma, Rahul Nagarajan, Lifeng Nai, Nishant Patil, Suvinay Subramanian, Andy Swing, Brian Towles, Cliff Young, Xiang Zhou, Zongwei Zhou, and David Patterson
Google, Mountain View, CA

{jouppi,gkurian,lsheng,pcma,rahulnagarajan,lnai,nishantpatil,suvinay,aswing,btowles,cliffy,zhoux,zongweiz}@google.com
and pattrsn@cs.berkeley.edu

## ABSTRACT

In response to innovations in machine learning (ML) models, production workloads changed radically and rapidly. TPU v4 is the fifth Google domain specific architecture (DSA) and its third supercomputer for such ML models. Optical circuit switches (OCSes) dynamically reconfigure its interconnect topology to improve scale, availability, utilization, modularity, deployment, security, power, and performance; users can pick a twisted 3D torus topology if desired. Much cheaper, lower power, and faster than Infiniband, OCSes and underlying optical components are <5% of system cost and <3% of system power. Each TPU v4 includes SparseCores, dataflow processors that accelerate models that rely on embeddings by 5x–7x yet use only 5% of die area and power. Deployed since 2020, TPU v4 outperforms TPU v3 by 2.1x and improves performance/Watt by 2.7x. The TPU v4 supercomputer is 4x larger at 4096 chips and thus nearly 10x faster overall, which along with OCS flexibility and availability allows a large language model to train at an average of ~60% of peak FLOPS/second. For similar sized systems, it is ~4.3x–4.5x faster than the Graphcore IPU Bow and is 1.2x–1.7x faster and uses 1.3x–1.9x less power than the Nvidia A100. TPU v4s inside the energy-optimized warehouse scale computers of Google Cloud use ~2-6x less energy and produce ~20x less CO2e than contemporary DSAs in typical on-premise data centers.

**ACM Reference Format:**
Norman P. Jouppi, George Kurian, Sheng Li, Peter Ma, Rahul Nagarajan, Lifeng Nai, Nishant Patil, Suvinay Subramanian, Andy Swing, Brian Towles, Cliff Young, Xiang Zhou, Zongwei Zhou, David Patterson. 2023. TPU v4: An Optically Reconfigurable Supercomputer for Machine Learning with Hardware Support for Embeddings: Industrial Product. In The 50th Annual International Symposium on Computer Architecture (ISCA '23), June 17–21, 2023, Orlando, FL, USA. ACM, New York, NY, USA, 14 pages. https://doi.org/10.1145/3579371.3589350.

## 1 INTRODUCTION

Happily for architects, machine learning (ML) models continue to evolve in challenging ways, both in scale and algorithmically (see Table 1 and Section 7.7). Examples of the former are *large language models* (*LLMs*) and examples of the latter are the embeddings necessary for recommender systems (*deep learning recommendation models* or *DLRMs*) and the huge calculations of Transformers and BERT. The incredible scale of recent LLMs [6, 38, 54] has stretched our ML supercomputer scale from 256 TPU v2 nodes to 4096 TPU v4 nodes. Reaching such a scale raises reliability problems that are particularly compounded by the HPC-style, checkpoint/restore, everything-must-work way that deep neural network (DNN) training is performed. It is far from the software reliability typical of mainline Google distributed systems.

This paper describes three major features of TPU v4 that respond to these challenges:

1. We addressed the scale and reliability obstacles by introducing *Optical Circuit Switches* (*OCSes*) with optical data links, allowing a 4K-node supercomputer through reconfiguration to tolerate 1K CPU hosts that are unavailable 0.1%–1.0% of the time.

2. We disclose the hardware support for embeddings in DLRMs (*SparseCore* or *SC*), part of TPUs since TPU v2.

3. Combining the first two capabilities, embeddings add all-to-all communication patterns to the demands on supercomputer-scale interconnect. Unlike all-reduce used in backpropagation, which maps well to 2D and 3D tori, all-to-all patterns strain bisection bandwidth. OCSes enable flexible topology configuration, including twisted torus [7], which has better bisection properties.

LLMs are a hot topic today in ML circles. While scale and reliability originally motivated OCSes in TPU v4, their topological flexibility and deployment advantages turned out to improve LLM training time significantly.

Since prior papers have described the fundamentals of previous TPUs for training [26, 39] and for inference [25, 27], this paper focuses on the three novel features of TPU v4 listed above that have not yet been described. The major contributions of the paper are:

- It describes and evaluates the first production deployment of OCSes in a supercomputer and the first to allow topology reconfiguration to improve performance.
- It describes and evaluates the first accelerator support for embeddings in a commercial ML system.
- It documents the rapid change in production model types since 2016 for the fast changing ML field (Table 1).
- It shows how Google uses ML to co-optimize DNN models, OCS topology, and the SparseCore.

The next section introduces OCSes and explains their many benefits. Section 3 motivates the SparseCore and shows its performance gains. Section 4 uses ML to search how to co-optimize the hardware and DNN models. The next two sections compare performance on production workloads versus TPU v3 and then versus the Nvidia A100 and the Graphcore MK2 IPU using MLPerf. The paper ends with a discussion, a related work section, and a summary.

**Table 1: Workloads by DNN model type (% TPUs used).** Over 90% of training at Google is on TPUs. The parenthesized entries split Transformer models into the subtypes of BERT and LLM. Columns 2 to 4 show workloads for inference [25], training and inference [26], and inference [27]. The last workload is for training on TPU v4s over 30 days in October 2022.

| DNN Model | TPU v1 7/2016 (Inference) | TPU v3 4/2019 (Training & Inference) | TPU v4 Lite 2/2020 (Inference) | TPU v4 10/2022 (Training) |
|---|---|---|---|---|
| MLP/DLRM | 61% | 27% | 25% | 24% |
| RNN | 29% | 21% | 29% | 2% |
| CNN | 5% | 24% | 18% | 12% |
| Transformer | -- | 21% | 28% | 57% |
| (BERT) | -- | -- | (28%) | (26%) |
| (LLM) | -- | -- | -- | (31%) |

## 2 RECONFIGURABLE OPTICAL SWITCH

We wanted to scale up the number of chips by 4x versus TPU v3 just as TPU v3 was 4x TPU v2. Given the distance between TPU v3 racks, some wrap-around links of its 2D torus topology were so long that they had to be optical due to the reach limitation of electrical interconnects. Optical links are >10x more expensive than electrical links. At 4x the scale, there would be even more optical links. Moreover, there were concerns about the bisection bandwidth of a 2D torus of that size and the availability of a single system of that scale. Using a 3D torus increases bisection bandwidth and the OCS acts like a plugboard to skip failed units.

### 2.1 Optical Circuit Switching

To improve data center networking, Google advanced the state-of-the-art in reliability and cost of optical transceivers and OCSes [43,54]. The resulting Google *Palomar OCS* is based on 3D Micro-Electro-Mechanical Systems (MEMS) mirrors that switch in milliseconds. They employ circulators to send light both ways in a fiber, halving the number of required ports and cables.

What size of the electrically-cabled building block to use? Given the 3D torus, 3D cubes have the best bisection bandwidth, suggesting 4×4×4 (64 chips) or 8×8×8 (512). With 4 TPU v4s per CPU host, 64 TPU v4 chips and their 16 CPU hosts comfortably fit into one rack. As 512 chips need multiple racks, a $4^3$ building block was chosen.

### 2.2 Construction of the TPU v4 Supercomputer

Figure 1 shows the links from the 6 "faces" of a $4^3$ block. There are 16 links per face, totaling 96 optical links per block that connect to OCSes. To provide the wraparound links of a 3D torus, the links on the opposing sides must connect to the same OCS. Thus, each $4^3$ block connects to $6 \times 16 \div 2 = 48$ OCSes. The Palomar OCS is 136×136 (128 ports plus 8 spares for link testing and repairs), so 48 OCSes connect the 48 pairs of cables from 64 $4^3$ blocks (each 64 chips), yielding the desired total of 4096 TPU v4 chips.

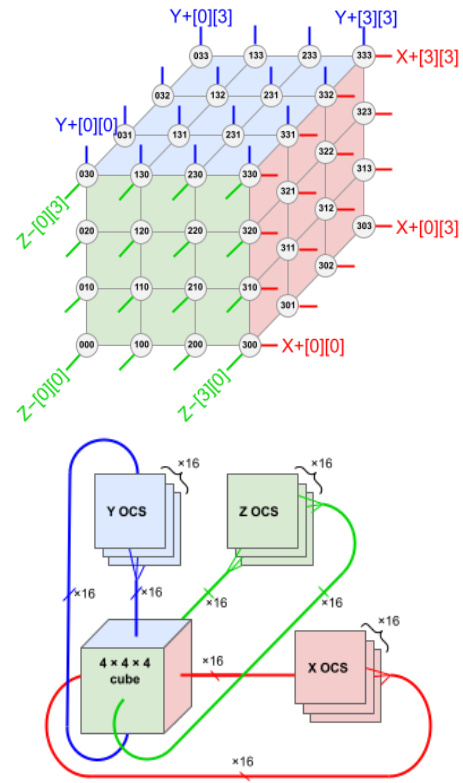

**Figure 1:** Connectivity of a 4×4×4 cube (top) to 3 OCSes (bottom). The "+" and "–" connections with the same dimension and index are connected to the same OCS; 48 of these in-out pairs each connect to a distinct OCS.

Figure 2 below shows a TPU v4 package and four of them mounted on the printed circuit board. Like TPU v3, each TPU v4 contains two *TensorCores* (*TC*). Each TC contains four 128x128 *Matrix Multiply Units* (*MXUs*) and a *Vector Processing Unit* (*VPU*) with 128 lanes (16 ALUs per lane) and a 16 MiB *Vector Memory* (*VMEM*). The two TCs share a 128 MiB *Common Memory* (*CMEM*). The PCB embeds 4 *Inter-Core Interconnect* (*ICI*) links, connected as a 2×2 mesh; 16 external ICI links go to other trays for constructing the 3D torus. Figure 3 below shows one row of eight racks, where each rack contains 16 tray-host server pairs. Passive electrical cables create a 4×4×4 3D mesh in a rack. Electrical-to-optical conversions happen at the fiber connector to the TPU trays. There are no other conversions until the light reaches an optical-to-electrical converter at the fiber connector of the destination tray. The 48 OCSes join eight rows together to form the complete 64-rack system.

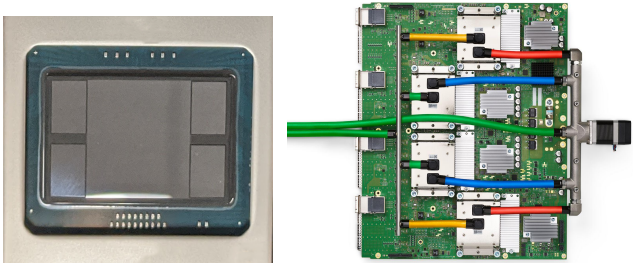

**Figure 2: The TPU v4 package (ASIC in center plus 4 HBM stacks) and printed circuit board with 4 liquid-cooled packages.** The board's front panel has 4 top-side PCIe connectors and 16 bottom-side OSFP connectors for inter-tray ICI links.

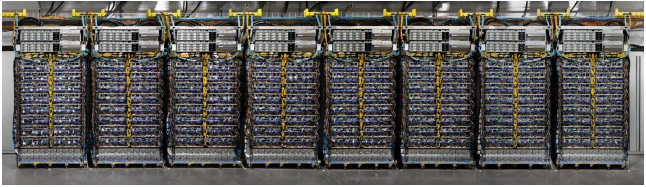

**Figure 3: Eight of 64 racks for one 4096-chip supercomputer.**

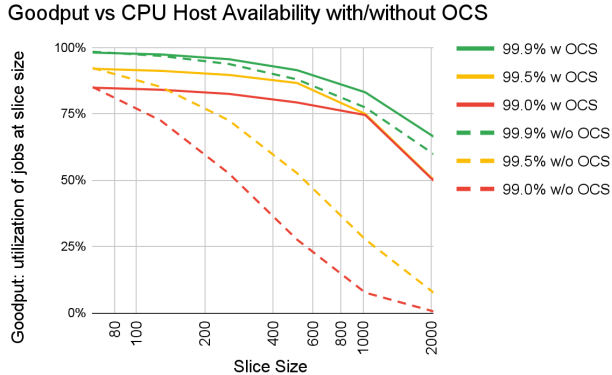

**Figure 4: Impact of OCS connected versus a statically connected supercomputer on goodput (i.e., effective throughput)** as CPU availability and slice size varies on a log scale. Goodput is counterintuitive at large slices. At ¼ of the 4K chips, goodput for both 99.0% and 99.5% is 75%, as 3 slices occupy ¾ of the chips. Spares are needed to allow scheduling jobs despite some failed nodes, so you can't realistically schedule two 2k node slices from 4k nodes. With one 2k node slice (50% of 4k), you have 50% spares, so it will have 50% goodput. With 3k nodes (75% of 4k), you have 25% spares, and therefore 75% goodput.

### 2.3 OCS Availability Benefits

An OCS raises availability by routing around failures. The main problem is the CPU host; each host has 4 TPU v4s, so 1K hosts per supercomputer. Like HPC supercomputers, the workload consists of a range of scale sizes, called *slices*: 64 chips, 128 chips, 256 chips, and so on. Figure 4 shows the "goodput" of slice sizes as host availability varies from 99.0% to 99.9% with and without OCSes. Without OCSes, host availability must be 99.9% to offer reasonable slice goodput. OCSes also have fair goodput for 99.0% and 99.5% for most slice sizes. Figure 4 assumes all slice size requests are equal, but workloads have many sizes (Table 2).

**Table 2: Sampling of popularity of TPU v4 slices for a day in November 2022.** This table includes all slices used ≥ 0.1%. Twistable (Section 2.8) but not twisted means the slice geometry allows twisting (n×n×2n or n×2n×2n), but the user picks the regular topology. The software scheduler requires that slices have dimensions x ≤ y ≤ z. Half of the slices have x, y, and z as either 4 or 8.

| Chips | <64 | | 64 | |
|---|---|---|---|---|
| Regular Tori | 1×1×1 (1) | 2.1% | 4×4×4 (64) | 13.9% |
| | 1×1×2 (2) | 0.4% | | |
| | 1×2×2 (4) | 6.7% | | |
| | 2×2×2 (8) | 4.7% | | |
| | 2×2×4 (16) | 6.4% | | |
| | 2×4×4 (32) | 8.9% | | |
| Total % | | 29% | | 14% |
| Chips | 128-192 | | 256-384 | |
| Twisted Tori | 4×4×8_T (128) | 16.0% | 4×8×8_T (256) | 9.2% |
| Twistable, not twisted Tori | 4×4×8_NT (128) | 1.5% | 4×8×8_NT (256) | 1.5% |
| Regular Tori | 4×4×12n (192) | 0.7% | 4×4×16 (256) | 1.0% |
| | | | 4×8×12 (384) | 0.1% |
| Total % | | 18% | | 12% |
| Chips | 512-768 | | 1024-1536 | |
| Twisted Tori | | | 8×8×16_T (1K) | 1.8% |
| Twistable, not twisted Tori | | | 8×8×16_NT (1K) | 1.4% |
| Regular Tori | | | 4×16×16 (1K) | 0.3% |
| | 8×8×8 (512) | 9.6% | 4×4×64 (1K) | 0.1% |
| | 4×8×16 (512) | 1.7% | 4×8×32 (1K) | 0.1% |
| | 4×4×32 (512) | 0.6% | 8×12×16 (1.5K) | 0.1% |
| | 8×8×12 (768) | 0.7% | 4×4×96 (1.5K) | 0.1% |
| | | | 8×8×24 (1.5K) | 0.1% |
| Total % | | 13% | | 4% |
| Chips | 2048-3072 | | | |
| Twisted Tori | 8×16×16_T (2K) | 1.4% | | |
| Twistable, not twisted Tori | 8×16×16_NT (2K) | 0.3% | | |
| Regular Tori | 12×16×16 (3K) | 5.7% | | |
| | 4×4×192 (3K) | 0.4% | | |
| Total % | | 8% | | |

### 2.4 OCS Deployment Benefits.

The OCSes also shrank deployment time. TPU v3 systems were not usable until all 1024 chips and all cables were installed and tested. Delivery delays for any component held up the entire supercomputer. For TPU v4, OCSes made each rack independent, so each $4^3$ block was put into production as soon as 64 chips and the necessary cables were installed and tested. Incremental deployment greatly improved the time to production use and thus cost effectiveness of the TPU v4 supercomputers.

## 2.5 OCS Scheduling Benefits

The OCS also simplifies scheduling, which increases utilization. For TPU v3, a 256 chip slice meant the scheduler had to find 256 contiguous chips that were idle. For TPU v4, it can pick four $4^3$ blocks from anywhere in the supercomputer. Slices don't even need to be a power of 2; they can be 4i×4j×4k, where 0 < i ≤ j ≤ k. For example, a user could request a 192 TPU v4 slice with a geometry of 4×4×12.

## 2.6 OCS Modularity and Security Benefits

Since the OCS can switch circuits in milliseconds, TPU v4 can easily change topology to match the application, the number of nodes, and the system that runs those jobs. TPU v4 provides wraparound links of a 3D Torus for most slice sizes, which doubles both the bisection bandwidth and the bandwidth of important collective communication operations (e.g., all-reduce) versus the mesh-like alternative [12], yet still allowing the TPU v4 to scale interconnect bandwidth up to $16^3$ (4096) chips. OCS also enables an air gapped network isolation between different slices, which enhances the security of multiple customers sharing a TPU v4 supercomputer.

## 2.7 Tailoring OCS Topology to Improve Performance

The final benefit was a bonus beyond solving problems of large scale. To set the stage, here are the three fundamental types of parallelism that improve the training time of DNNs:

   1. *Data Parallelism*: Each chip computes the forward and backward pass on a subset of examples, and sends the gradients that it calculates for its subset to the other chips.

   2. *Model (or Tensor) Parallelism*: Large tensor operations and their weights are divided across multiple chips, so that each chip simultaneously computes a subset of a tensor operation.

   3. *Pipeline Parallelism*: For a DNN with many layers, each chip computes a subset of layers, and communicates the layer results to chips holding the adjacent layers.

   Users can change the TPU v4 topology to match the type of parallelism being used, e.g., for a 512 slice, pipeline parallelism might want a cigar shape (4×4×32) instead of the conventional $8^3$ cube (8×8×8). For the highest bisection bandwidth, often needed by embedding heavy applications, the conventional $8^3$ cube is preferred. ML practitioners often combine parallelism types to get the best performance, such as data plus model parallelism. Model parallelism typically has two parameters: width and length. To take full advantage of the bandwidth available, users map data parallelism along one dimension of the 3D torus and the two model parallel parameters on the other dimensions. Table 3 in Section 4 gives examples of performance gains of 1.2x to 2.3x by varying the topology and the hyperparameters.

   TPU v4 uses a single, static topology for each training job, which can be co-optimized with the communication requirements of the training job. This per-job configuration is not a fundamental limitation of the OCS. See [55] for more details of the OCS and its physical construction.

## 2.8 Twisting the Torus

For a slice with a number of TPUs that is a perfect cube, a symmetric torus with an equal number of TPUs along each dimension minimizes latency and maximizes bisection bandwidth (e.g., $8^3$ = 512 TPUs organized as an 8x8x8 torus). For non-perfect-cube slices, a rectangular torus can be built with different numbers of TPUs along each dimension. Alternatively, [7] proposes a topology that outperforms rectangular tori with lower latency and higher bisection bandwidth without increasing switch hardware. The *twisted torus* rewires some links between $4^3$ cubes to reduce the worst case latency. Figure 5 shows a regular topology and a twisted topology. Since TPU v4 uses an OCS to connect $4^3$ blocks, the "rewiring" is mostly reprogramming of routing in the OCS. Using all-to-all communication with large messages as a microbenchmark, Figure 6 below shows the performance gain from the twisted topology. The twisted torus improves all-to-all throughput by 1.63x and 1.31x over the regular torus on 4×4×8 and 4×8×8 slices, respectively. While the popular interconnect topologies today are Clos networks and Dragonfly networks, the twisting option reduces the worst case bisection bandwidth for the 3D tori and makes them more attractive for today's supercomputers.

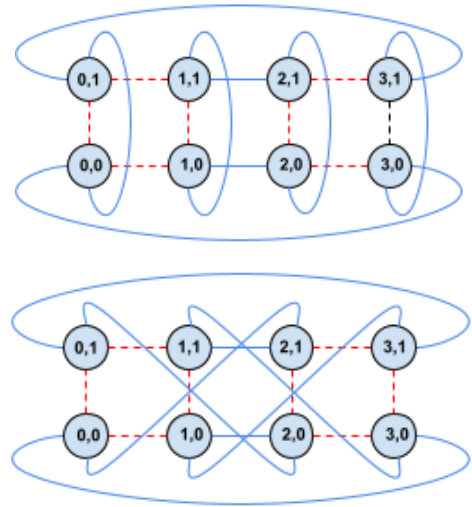

**Figure 5: Example of regular (top) and twisted torus (bottom) topologies for a 4×2 slice of TPU v4 nodes.** The TPU v4 network is three-dimensional, but the figure uses two dimensions for ease of illustration. Each TPU is labeled with its coordinates in the slice. The electrical connections (red dashed lines) remain fixed. By utilizing the flexibility of the OCSs, the optical connections (blue solid lines) can be reconfigured from a rectangular torus to a twisted torus without any physical recabling of the machine; the only change is in the routing tables. TPU v4 uses a k×k×2k configuration from Camarero, Martinez, and Beivide [8].

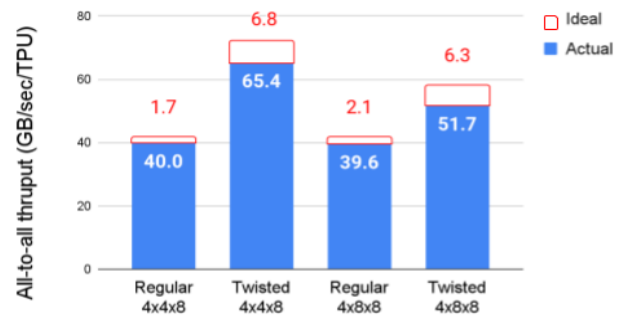

**Figure 6: Measured all-to-all throughput for 4×4×8 and 4×8×8 slices using regular and twisted tori.** Measurements are steady state (large aggregate transfer size) with individual DMAs being 4 KiB. Each column also shows the theoretical delta from the ideal peak as a stacked bar and label above the measured performance.

## 2.9 Distribution of Topologies

Table 2 above gives a sampling of slice topologies used in production. Some tasks are smaller than a $4^3$ block, so they can only use a 2D mesh; 29% are smaller than a $4^3$ cube, so they obviously cannot select a twisted 3D torus. Of the remaining 71%, only those of the form n×n×2n or n×2n×2n (n≥4) can twist. They are 33% (48% of 71%). The actual twisted tori are 28% (86% of 33%). Stated alternatively, 40% of the topologies that are $4^3$ blocks or larger use twisted tori.

## 2.10 Cost of OCS Flexibility

Remarkably, given all the benefits of OCSes, their cost is <5% of the total TPU v4 supercomputer capital costs and <3% of total power. The power and cost accounting includes the entire optical fabric, including the optics modules, fiber, and OCS infrastructure.

# 3. SPARSECORE: EMBEDDINGS SUPPORT

Before introducing the next TPU innovation, let's review recommendation models, embeddings, and distributed training to set the stage.

## 3.1 Recommendation Models

Deep learning recommendation models (DLRMs) are a quarter of our ML workload (Table 1 above). DLRMs are used in advertising, search ranking, YouTube, and Google Play applications [1, 4, 11, 13]. Google's production advertising models score ads for billions of queries daily, and consist of billions of weights, and train on more than one trillion examples, and are required to perform inference at well over one hundred thousand requests per second [1]. DLRM model sizes are determined using five factors: prediction quality, training time, total training cost, serving latency, and total serving cost. Embeddings are a key component of DLRMs.

## 3.2 Embeddings

Inputs for DLRMs consist mainly of categorical features. Each categorical feature contains N discrete values (N is commonly referred to as the vocabulary size). For example, in the search ranking application, the search query is a categorical feature, and N is the number of words in the English language. A given training example (query) is sparse, and contains a tiny subset of words.

Neural networks typically train well on dense vectors. Embeddings are the standard and effective way to transform categorical feature values into dense vectors. An embedding function translates from the large, categorical space (e.g., words in the English language) to a smaller, dense space (e.g., a 100-vector representing each word). Embedding functions are implemented using lookup tables. An example is a table with 80,000 rows (one per word) of width 100. Each training example can look up either one row (univalent embedding) or a small, dynamic number of rows (multivalent embedding, typically combined by summing). A neural network model might have many tables of many sizes for different categorical features. Embeddings are a key component of Google DLRMs, and typically form the first layer in a neural network model.

## 3.3 Distributed Training

Embedding tables are large, and can range in size from O(10 MiB) to O(100 GiB). In aggregate, all the embedding tables in a model can be as large as several TiBs. Hence, such tables are partitioned across the memory of several TPU chips. There are three methods for partitioning: (1) *column sharding* splits tables along their width across multiple chips, (2) *row sharding* splits tables along their vocabulary size, and (3) *table sharding* places different tables on different chips. These distribution strategies are collectively termed model parallelism in the context of neural network models. For small embedding tables, replication across all chips (using data parallelism) is better for performance.

## 3.4 Key Performance Attributes

Embedding lookup operations consist mainly of small gather or scatter memory accesses, which have low arithmetic intensity. As opposed to dense operations (e.g., transformers, fully connected networks) where the chip FLOPS/second is the main driver of end-to-end performance, embedding lookup operations are bottlenecked by the memory bandwidth, memory capacity, and VPU (vector processing unit) performance. The ICI interconnection network (across chips) is also a significant performance driver.

The interconnect bandwidth and performance depends on the type of parallelism being exploited. For model parallelism (common case), the communication pattern consists of variable-length all-to-all exchange. The network bisection bandwidth limits performance. For data parallelism, the communication pattern consists of all-reduce operations, which injection bandwidth limits.

The unstructured sparsity of embeddings is also prone to compute, memory, and communication load imbalances across a supercomputer. To reduce load imbalance, deduplication of frequent feature values is commonly used, and must be efficiently supported by the compute substrate. Deduplication also reduces the number of memory accesses, and the quantity of data sent over the interconnection network, further improving performance.

## 3.5 SparseCore

It's time to introduce the *SparseCore* (*SC*). For the training phase, embeddings could be placed on the TensorCore or the host CPUs of the supercomputer. The TensorCore has wide VPU and matrix units, and is optimized for dense operations. Placing embeddings on the TensorCore would be suboptimal due to small gather/scatter memory accesses, and variable length data exchange. Placing embeddings on the host CPUs of a supercomputer would induce an Amdahl's Law bottleneck over the CPU DRAM interface, amplified by the 4:1 TPU v4 to CPU host ratio. Tail latency and bandwidth restrictions of data-center networking would further constrain the training system.

Performance could be optimized using the total HBM capacity of a TPU supercomputer, joined by a dedicated ICI network, and with fast gather/scatter memory access support. This insight led to the codesign of the SparseCore (SC).

The SC is a domain-specific architecture for embedding training starting with TPU v2, with later improvements in TPU v3 and TPU v4. SCs are relatively inexpensive, at a total of only ~5% of the die area and ~5% of the power. SCs operate in a sea-of-cores configuration, combining supercomputer-scale HBM and ICI to create a flat, globally addressable memory space (128 TiB in TPU

v4). In contrast to all-reduces of large parameter tensors in dense training, the all-to-all transfers of smaller embedding vectors use HBM and ICI with finer-grained access patterns for scatter/gather.

As separate cores, SCs allow parallelization across dense compute, SC, and ICI communications. Figure 7 shows the SC block diagram, which we consider a "dataflow" architecture because data flows from memory to a variety of directly connected specialized compute units.

The most general SC units are the 16 compute tiles (dark blue boxes in Figure 7). Each tile has an associated HBM channel and supports multiple outstanding memory accesses. Each tile has a *Fetch Unit*, a programmable 8-wide SIMD *Vector Processing Unit* (*scVPU*, not to be confused with VPU of the TC in TPU v4), and a *Flush Unit*. The Fetch Unit reads activations and parameters from the HBM into the tile's slice of a 2.5 MiB *Sparse Vector Memory (Spmem)*. The scVPU uses the same ALUs as TC's VPU. The Flush Unit writes updated parameters to HBM during the backward pass. In addition, the five *Cross-Channel Units* (gold boxes in Figure 7) perform specific embedding operations, which their names explain. Like TPU v1, the units execute CISC-like instructions and operate on variable-length inputs, where the run-time of each instruction is data-dependent. The cross-channel units operate across all 16 banks of Spmem collectively.

### 3.6 SparseCore Performance

The end-to-end embedding lookup performance is essentially proportional to the bisection bandwidth due to the all-to-all transfers of small embedding vectors. For the 2D torus used in TPU v2 and TPU v3, this bandwidth scales as $N^{1/2}$ for N chips. The 3D torus in TPU v4 scales as $N^{2/3}$ [12]. Figure 8 shows that the TPU v3/v4 bisection bandwidth ratio is 2–4x higher at a given chip count and accelerates embeddings by 1.1x–2.0x. At 1024 chips, SC overheads start to dominate, so bisection bandwidth is less important.

Figure 9 below shows performance of an internal production recommendation model (DLRM0, see Sections 7.8 and 7.9) across the two TPU generations for 128 chips. The standalone CPU configuration has 576 Skylake sockets (400 for learners and 176 for variable servers). The bottom two bars show TPU v4 without SC, where the embeddings are placed in CPU memory. The "Emb on CPU" bar places embeddings in CPU host memory and the "Emb on Variable Server" bar places embeddings on 64 external variable servers. TPU v3 is faster than CPUs by 9.8x. TPU v4 beats TPU v3 by 3.1x and CPUs by 30.1x. When embeddings are placed in CPU memory for TPU v4, performance drops by 5x–7x, with bottlenecks due to CPU memory bandwidth.

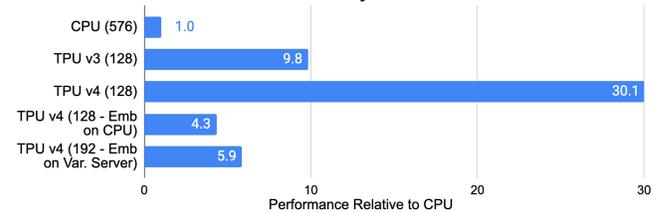

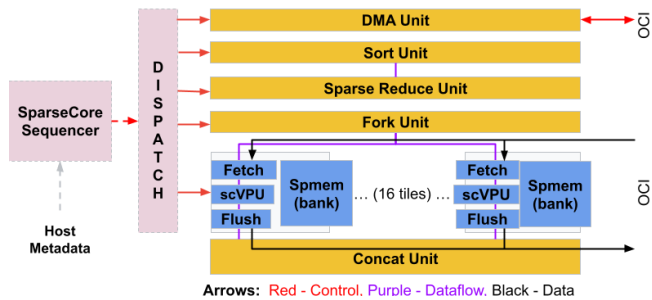

**Figure 7: SparseCore (SC) Hardware Architecture.**

**Figure 9: Performance of an internal recommendation model (DLRM0) on CPUs,** TPU v3, TPU v4, and TPU v4 with embeddings in CPU memory (not using SparseCore). The numbers in parenthesis indicate the number of sockets (CPU or TPU) used for training.

## 4  USING ML TO TAILOR THE DNN TO THE TPU AND THE TPU TOPOLOGY TO THE DNN

To enable Pareto-optimizations over quality and performance for DNN models, we developed platform-aware neural architecture search (PA-NAS) at scale to tailor DNN models for TPU v4 supercomputers automatically [32]. A PA-NAS designed CNN1 achieves ~1.6X better performance (QPS and latency) than the baseline designed by generic NAS, with comparable accuracy [32]. Unlike [33], here we show how PA-NAS improved the performance of DLRM0 on TPU v4.

DLRMs are twice as popular as CNNs for our ML workloads (see Table 1 above). Unlike CNNs, which mostly use TensorCores (TCs), DLRMs use both SCs and TCs. PA-NAS can shift computation load between sparse layers (running on SCs) and dense layers (running on TCs) for Pareto-optimal performance and quality.

Figure 10 below shows PA-NAS on a production-scale DLRM model. Despite having been optimized manually and by using a generic NAS, the original DLRM0 idled the SC ~25% of the execution time (top blue bar in Figure 10) because of the load imbalance between SCs and TCs. PA-NAS enables the end-to-end Pareto-optimizations on quality and performance over both embedding layers (running on SC) and hidden layers (running on TC) for DLRM0, which approaches perfect SC-TC load-balance (lower blue and red bars in Figure 10) and improves DLRM0 end-to-end performance by >10%. This performance uplift is equivalent to improvements historically achieved by a team of >10 experts over about half a year, further demonstrating PA-NAS's capability of increasing accelerator flexibility and performance gain.

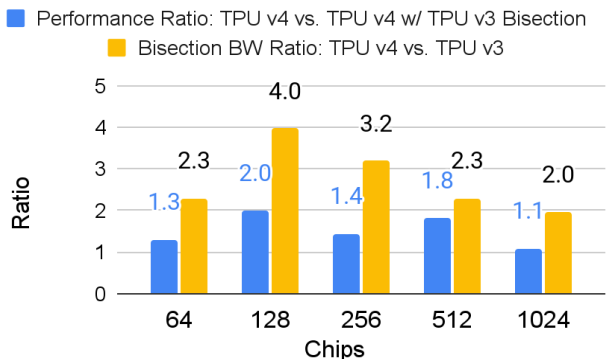

**Figure 8: Bisection bandwidth ratio of TPU v4 to TPU v3 and performance sensitivity to bisection bandwidth.** The model used is a DLRM with ~100M dense parameters in fully connected layers, ~20B embedding parameters (~300 features mapped to ~150 tables), and 1-100 average valency per feature. The global batch size is scaled proportionately to the number of chips.

We can also use search to tailor the TPU v4 topology to the DNN Model. Table 3 shows gains in performance from searching for configurations for two models. The first example increased performance 2.3x by changing the geometry for a 512 TPU v4 slice over the novice user's initial design for an LLM. The second example shows a harder task, demonstrating an improvement of 1.2x over an expert's design for the pre-training phase of GPT-3.

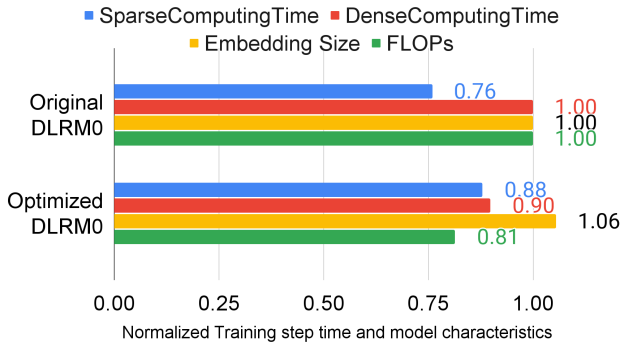

**Figure 10: Performance improvements by PA-NAS for DLRM0.** Since DLRMs use both SCs and TCs, the maximum of SparseComputingTime and DenseComputingTime is the end-to-end training step time of a DLRM. All computing times in the figure (SparseComputingTime and DenseComputingTime for both original and optimized DLRM0) are normalized against the DenseComputingTime of the original DLRM0, since the original DLRM0 is bottlenecked on Dense Computing. Embedding sizes and FLOPs are normalized against those of the original DLRM0.

**Table 3: Improvements in performance as we vary the topology of a 512 TPU v4 slice for training an LLM and the GPT-3 pre-training stage.** The original LLM used model parallelism of dimensions 16×32 and no pipeline or data parallelism for a 4×8×16 topology. The revision changed model parallelism to 64×8 and the topology to 8×8×8. The "1D/2D activation/weight partitioning" option is a typical way of partitioning tensors in a large model graph (see Figure 7 in [63]). For GPT-3, the original row used a 8×8×8 topology, pipeline parallelism of depth 8, no data parallelism, and model parallelism of dimensions 8×8. The revision changed the topology to 4×8×16, pipeline depth to 16, data parallelism to 4, model parallelism parameters to 1×8.

| Case | Versions | Through-put (seqs/sec) | Hyper-Parameters (topology, partition spec [pipeline, data, model1, model2], 1D/2D activation/weight partitioning) |
|---|---|---|---|
| LLM | Novice's pick | 17.9 (1.0x) | 4×8×16, [1, 1, 16, 32], 2D/2D |
| | Best perf. | 41.3 (2.3x) | 8×8×8, [1, 1, 64, 8], 1D/2D |
| GPT-3 Pre-training | Expert's pick | 21.0 (1.0x) | 8×8×8, [8, 1, 8, 8], 2D/2D |
| | Best perf. | 25.0 (1.2x) | 4×8×16, [16, 4, 1, 8], 1D/1D |

## 5 PRODUCTION WORKLOAD PERFORMANCE

Table 1 above shows the workload mix for 2022 plus the history of how it has changed over time; Section 7.7 discusses the workload changes. We use 8 applications to capture the production workload to compare TPU v3 to TPU v4. Figure 11 shows how efficiently eight production workloads scale up on TPU v4. We are encouraged that half of the workloads (CNN0, RNN0, RNN1, and BERT1) scale well to 3K chips. (Figure 4 above shows that given CPU hosts with availability of 99.0% to 99.9%, in practice it is much easier to schedule a 3K slice than a 4K slice.) For the remainder, our production teams are aware of where the scaling limits are and are constructing solutions, but those solutions are not yet implemented to allow measuring performance at full 3K scale

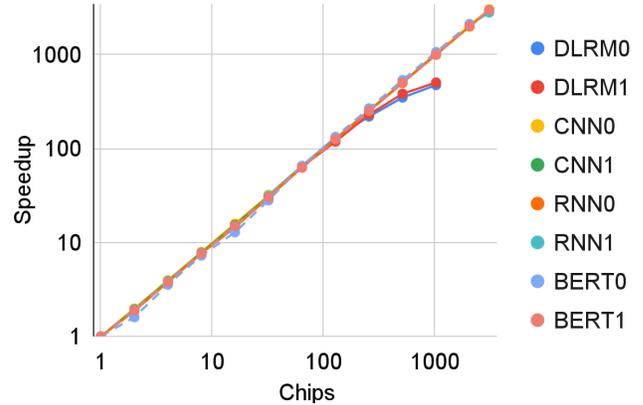

**Figure 11: Scalability of TPU v4 production workloads on a log-log scale.** Infrastructural limitations currently hinder getting the last few data points: BERT0 scales to 2K, DLRM0/1 to 1K.

**Table 4: TPU v4 and TPU v3 [26] features. Measured power is for the ASIC and HBM running production applications.**

|  | Google TPUv4 | TPUv3 |
|---|---|---|
| Production deployment | 2020 | 2018 |
| Peak TFLOPS | 275 (bf16 or int8) | 123 (bf16) |
| Clock Rate | 1050 MHz | 940 MHz |
| Tech. node, Die size | 7 nm, <600 mm2 | 16 nm, < 700 mm2 |
| Transistor count | 22 billion | 10 billion |
| Chips per CPU host | 4 | 8 |
| TDP | N.A. | N.A. |
| Idle, min/mean/max power | 90, 121/170/192 W | 123, 175/220/262 W |
| Inter Chip Interconnect | 6 links @ 50 GB/s | 4 links @ 70 GB/s |
| Largest scale configuration | 4096 chips | 1024 chips |
| Processor Style | Single Instruction 2D Data | Single Instruction 2D Data |
| Processors / Chip | 2 | 2 |
| Threads / Core | 1 | 1 |
| SparseCores / Chip | 4 | 2 |
| On Chip Memory | 128 (CMEM) + 32 MiB (VMEM) + 10 MiB (spMEM) | 32 MiB (VMEM) + 5 MiB (spMEM) |
| Register File Size | 0.25 MiB | 0.25 MiB |
| HBM2 capacity, BW | 32 GiB, 1200 GB/s | 32 GiB, 900 GB/s |

Table 4 compares key features of TPU v3 and TPU v4. Manufactured in 7 nm instead of 16 nm, TPU v4 has twice the matrix multipliers (enabled by the increased process density) and an 11% faster clock—this drives the 2.2X gain in peak performance. About 40% of the performance/Watt improvement was from technology and the rest was from design improvements (e.g., balancing the pipeline, implementing clock gating). The HBM memory bandwidth is 1.3x higher. Depending on the slice size, the bisection bandwidth of TPU v4 is 2x–4x (see Figure 8 above). It also has the 128 MB on-chip CMEM scratchpad memory not found in TPU v3.

Figure 12 shows how much faster TPU v4 supercomputers are than TPU v3 supercomputers at the same slice size. Given the comparisons in Table 4, it's not surprising that at the same slice size most applications run 1.5x-2.0x faster on TPU v4 than on TPU v3. DLRM0 is 3.0-3.5x faster and DLRM1 is 2.8x at 512 chips as TPU v4 has twice as many SCs and their clock rate is faster. The surprise is RNN1; it runs 3.3x faster on TPU v4. RNN1's small weights and small batch size benefit significantly from CMEM bandwidth versus HBM.

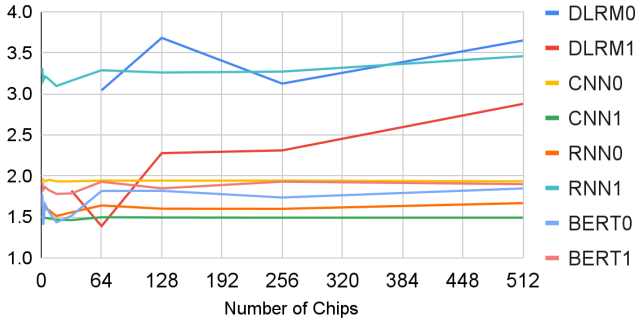

**Figure 12: Speedup of TPU v4 vs v3 for the same slice sizes.**

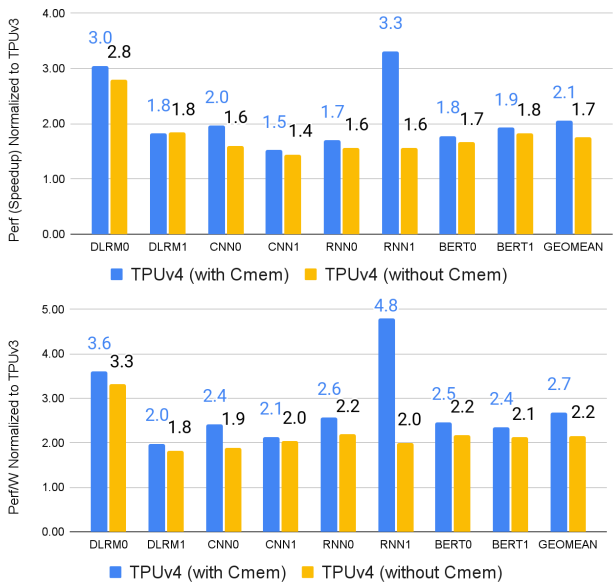

**Figure 13: Per chip performance (top) and package-level performance/Watt (bottom) for production applications relative to TPU v3 for CMEM turned on and off for smaller systems (e.g, 32 chips).** DLRM1 is much faster at 512 chips. DLRMs here are different from MLPerf DLRM (see Section 7.9).

Figure 13 shows results with CMEM turned off on TPU v4; it contributes to 1.2x performance gain overall but 2x for RNN1. It also shows that TPU v4 has 2.1x the performance and 2.7x the performance/Watt of TPU v3; as mentioned above, ~40% of the gain was from the technology and the rest from design.

LLM training will become a benchmark in a future MLPerf release. We omit performance of internal LLMs—31% of the workload in Table 1—on TPU v3 because it is unoptimized. TPUv3's 2D fixed topology hinders high-performance model partitioning needed for LLMs. We also lack TPUv3 chip capacity to train large LLMs within reasonable time-to-convergence SLOs given their lower FLOPS/second and suboptimal model partitioning performance.

**Table 5: Features of the two DSAs [21, 40] reporting MLPerf 2.0 Training results besides TPU v4.** The A100 has 32×108 = 3456 threads and the IPU has 6×1472 = 8832 threads.

|  | **Nvidia A100** | **Graphcore MK2 IPU** |
|---|---|---|
| Production deployment | 2020 | 2021 |
| Peak TFLOPS | 312 (bf16), 624 (i8) | 250 (bf16) |
| Clock Rate Base/Boost | 1095 /1410 MHz | 1850 MHz |
| Tech. node, Die size | 7 nm, 826 mm$^2$ | 7 nm, 832 mm$^2$ |
| Transistor count | 54 billion | 59 billion |
| Chips per CPU host | 4 | 4 |
| TDP | 400 W | 300 W |
| Inter Chip Interconnect | 12 links @ 25 GB/s | 3 links @ 64 GB/s |
| Largest scale MLPerf 2.0 configuration | 4216 chips | 256 chips |
| Processor Style | Single Instruction Multiple Threads | Multiple Instruction Multiple Data |
| Processors / Chip | 108 | 1472 |
| Threads / Core | 32 | 6 |
| On Chip Memory | 40 MiB | 900 MiB |
| Register File Size | 27 MiB | 1.40 MiB |
| HBM2 capacity, BW | 80 GiB, 2039 GB/s | 0 |

## 6 MLPERF BENCHMARK PERFORMANCE

This section compares DSAs based on the already published results for MLPerf Training [36]. Table 5 compares the features of two of the entries: the NVIDIA A100 and Graphcore MK2 IPU. While the top six rows are quite similar to TPU v4 in Table 4, the remaining rows show the diverse choices of the architects in terms of processor style, number of processors per chip, number of threads per processor, register file size, and on-chip vs. off-chip memory. First, TPU v4 has only 2 threads while A100 has 32×108 = 3456 threads and 6×1472 = 8832 threads for the IPU. The striking features of the IPU Bow are the ~1500 cores, *900 MB* of on-chip SRAM, and *zero* attached HBM or other DRAM. The striking features of the A100 are a 27 MB register file (to support multithreading) and only 40 MB of on-chip SRAM. Both chips use full-reticle dies, making them ~40% larger than the TPU v4 die.

Figure 14 below shows the fastest performance per DSA for five MLPerf benchmarks. (Graphcore ran two of the five.) Vendors are free to pick the size of the system for which they want to report results. Ideally, MLPerf would benchmark systems of equal size or cost or power, but that is not required.

Figure 15 below shows the reported results for ResNet and BERT as large points while the dashed lines between the points are interpolations based on the number of chips. The published MLPerf results for TPU v4 and A100 both scale to much larger systems than the IPU (4096 vs 256 chips). For similar sized systems, TPU v4 is 1.15x faster for BERT than the A100 and ~4.3x faster than the IPU. For ResNet, TPU v4 is 1.67x and ~4.5x faster, respectively.

Table 6 shows the power we measured running MLPerf benchmarks; A100s use on average 1.3x–1.9x more power.

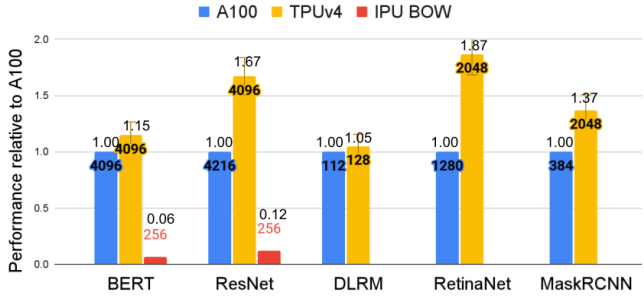

Figure 14: Reported MLPerf Training 2.0 highest performance [36] relative to A100. Each column labels the number of chips per system. Graphcore submitted results for BERT and ResNet. TPU v4 DLRM is in the research category. The MLPerf DLRM is not representative of production DLRMs [64] (see Section 7.9).

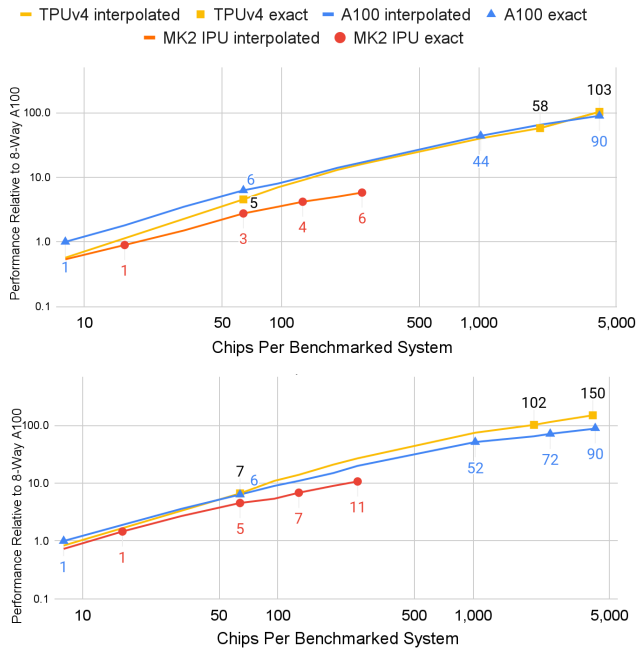

Figure 15: Reported MLPerf training 2.0 performance for BERT (top) and ResNet (bottom) [36] relative to an 8-Way A100 GPU on a log-log scale. To include IPUs, we only show BERT and ResNet in this figure. At the largest scale of 4096 chips, TPU v4 is 1.15x as fast as the Nvidia A100 for BERT. At 256 chips, the maximum IPU size in MLPerf, TPU v4 is ~4.3x as fast as the MK2 IPU Bow. At the largest scale, 4096 TPU v4s are 1.67x as fast as 4216 Nvidia A100s for ResNet. At 256 chips, TPU v4 is ~4.5x as fast as the MK2 IPU Bow. The points are the reported results, and the dashed lines are interpolations for intermediate sizes systems. For TPU v4, the results for ≤2048 chips are from MLPerf Training 1.0; all the other points for all systems are from MLPerf Training 2.0.

Table 6: Mean power for DSA plus HBM for 64 chip systems running MLPerf. Adding the switches would show even higher efficiency for TPU v4 (Section 7.3). We use nvidia-smi reported power measurement on Azure Standard_ND96amsr_A100_v4 VMs during a rerun of Nvidia's MLPerf 2.0 64-chip submission. TPU v4 power measurements are done by running Google MLPerf 2.0 benchmark code on 64-chip scale in Google data center. The TPU v4 mean power measurement is 2%–8% higher than in Table 4, but the workloads differ. MLPerf 3.0 may add power measurements to performance in the October 2023 round.

| MLPerf Benchmark | A100  | TPU v4 | Ratio |
|------------------|-------|--------|-------|
| BERT             | 380 W | 197 W  | 1.93  |
| ResNet           | 273 W | 206 W  | 1.33  |

## 7 DISCUSSION

We comment next on 11 questions readers might have from our analysis of TPU v4 and the other DSAs.

### 7.1 Do peak FLOPS/second predict real performance?

Many in the ML community think peak FLOPS/second are a good performance proxy [45], but they are not. For example, TPU v4 is 4.3x–4.5x faster on two MLPerf benchmarks than IPU Bow on equal sized systems despite only having a 1.10x edge in peak FLOPS/second. Another example is that the A100 peak FLOPS/second rate is 1.13x TPU v4, but TPU v4 is 1.15x–1.67x faster for the same number of chips. Figure 16 gives the relationship between peak FLOPS/sec and memory bandwidth using the roofline model [61].

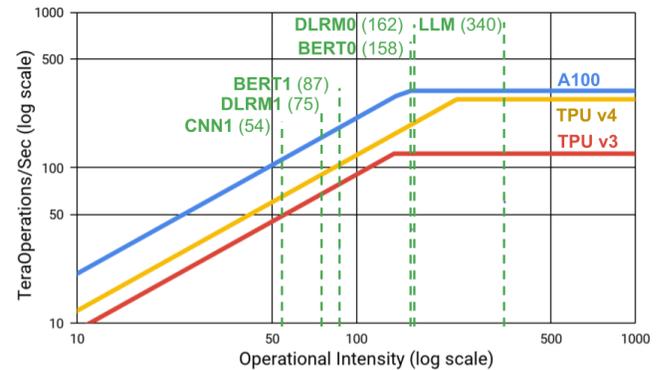

Figure 16: Roofline models for TPU v4, TPU v3, and A100 plus DNN models [61]. Operational intensity is in parentheses.

The A100 higher peak performance is for Boost Mode of up to 1410 MHz; the base clock is 1095 MHz. If the average rate was 1243 MHz, the peak performance of the A100 and TPU v4 would be equal (same ceiling in the roofline model). We measured the A100 running MLPerf BERT, and the average clock rate was 1280 MHz due to capping of power use by Nvidia GPU software.

Amdahl's Law reminds us that system balance—in compute, memory, and interconnect—with sufficient power and cooling to keep everything executing is still important. The integrated ICI network lets TPU supercomputers scale performance gracefully and the OCSes let users tailor the topology to the application to improve performance.

### 7.2 How does OCS differ from NVLink and NVSwitch?

Every TPU since 2017 has had its own built-in router between links (ICI) to neighboring chips in the torus. Like NVlink, it enables "glueless" connection of TPUs, but at a larger scale than 4-8 GPUs: 256 TPU v2s and 1024 TPU v3s.

We think of optical circuit switching as the next generation of ICI versus a response to the latest NVSwitch, which uses electrical packet switching for 8 GPUs with one switch and up to 256 GPUs with two levels of switches.

OCSes are just fibers connected by mirrors, so any bandwidth running through a fiber can be switched between input and output fibers by the OCS across 4096 chips today (or even more in the future). For example, an OCS could handle multiple terabits/second per link by using wavelength multiplexing. Moreover, all inputs can be connected to all outputs, but the connections must be 1:1.

### 7.3 What if TPU v4 used IB versus OCS?

Let's start with Infiniband (IB) versus OCS switches. Just as NVLink connects 8 GPUs in a DGX, 8 TPUs would use ICI. We follow Nvidia's guidance by using a full 3-level fat tree for the hybrid IB/ICI network [41]. At an average of one NIC per GPU, a 1120 A100 superpod needs 164 Mellanox QM8790 40-port IB switches [41], each priced at ~$15k–$18k [10, 23]. The 1120 IB NICs are extra. To replace the 48 128-port OCSes, 4096 TPU v4s need 568 IB switches. An OCS is no more expensive per port than an IB switch, but it can support higher bandwidth because it is passively reflecting light encoded at the source. The hybrid IB/ICI option is substantially more expensive *and* harder for software.

Furthermore, active packet processing for an IB switch is far more power hungry than the tiny amount of power required to hold the MEMS mirrors to their configured orientation in an OCS.

ICI link bandwidth is 2x IB—400 vs 200 Gbit/s—but system speed is harder to judge. An internal event-driven simulator that operates at the TensorFlow graph operation level evaluated a hybrid ICI/IB network. (It ignores protocol processing on the CPU, which can be significant.) Depending on the slice size, an optimized all-reduce would run 1.8x–2.4x slower and an all-to-all would be 1.2x–2.4x slower. This network heterogeneity is also a software challenge. As communication is only a portion of training time, overall IB slowdown for a DNN might be as little as ~10%.

However, the biggest impact is losing the benefits that originally inspired the use of OCS (Section 2): availability, scale, utilization, modularity, power efficiency, deployability, and so on.

### 7.4 Nvidia announced the H100, the successor to A100, in 2022. Why not compare TPU v4 to it?

After systems are running production applications in the field, the Google tradition is to write retrospective, peer-reviewed papers for prominent conferences. The rationale is that the intellectual incentives and deadlines for a prestigious publication encourage architects working on the next generation to take the time to reflect and to make careful, detailed, apples-to-apples comparisons to the previous chip and contemporary alternatives that can pass peer review. The lessons learned improve future designs. The good news is that these retrospective, peer-reviewed, apples-to-apples papers are widely read, e.g., [25] has >4000 citations. If Google sold chips versus using them internally, we might instead need to publish unreviewed whitepapers much earlier in the chip lifecycle.

Speaking of apples-to-apples, both TPU v4s and A100s were deployed in 2020 and both use 7 nm technology. The newer, 700W H100s were not available at AWS, Azure, or Google Cloud when we did the research in 2022 or even when we submitted the final camera ready paper in 2023. The appropriate H100 match would be a successor to TPU v4 widely deployed in a similar time frame and technology (e.g., in 2023 and 4 nm).

### 7.5 Why 30%–90% more power for A100 (Table 6)?

It is hard to find a complete quantitative answer for these two complex designs. The 4x larger on-chip SRAM (160 MB versus 40 MB) allows memory transfers to DRAM to be in larger blocks, improving energy efficiency. Figure 13 above shows turning on the CMEM local memory, which increases on-chip SRAM from 32 MB to 160 MB, improves performance by 1.18x and performance/Watt by 1.24x.

The following three qualitative factors could explain the rest of the gap, but we can't say definitively without additional work. Support for multithreading on the GPU leads to a 100x larger register file (27 MiB versus 0.25 MiB), which likely requires more energy for register accesses—even though the GPU uses a single port SRAM—as power generally increases with the square root of memory capacity [22]. Second, the 128x128 MXUs of TPU v4 mean each 128 entry input gets reused 128 times, whereas the 4x4 FP16 array multipliers of the A100 only get reused 4 times, leading to more on-chip SRAM accesses. Finally, the ~40% larger A100 chip may have longer data buses, increasing data transmission energy.

### 7.6 What is the CO2e from TPU v4 vs other DSAs?

There is considerable concern about the carbon footprint of ML [42, 45]. Let's compare the cloud-only TPU v4 to a hypothetical recent DSA in an on-premise data center.

Practitioners can reduce operational energy use and CO2 emissions by optimizing the "4Ms" [42]:

1. Let's assume TPU v4 and another DSA are training the same models, so the *Model* parameter is 1.0 in this case.

2. The *Machine* parameter is measured in performance/Watt. TPU v4 is 2.7x TPU v3. The MLPerf power plan is in progress, so we estimate for others. TPU v4 is 1.2x–1.7x faster and 1.3x–1.9x lower power than the A100 and 4.3x-4.5x faster than the IPU, whose TDP is 300 W. TPU v4 chip performance/Watt is thus ~2x–6x versus a contemporary DSA; to be conservative, we assume 2x for this calculation.

3. The *Mechanization* parameter is data center power efficiency, measured as *Power Usage Effectiveness* (*PUE*). For on premise data centers, PUE is often high. Computer architects improved PUE by helping advance the state-of-the-art of *warehouse scale computers* (*WSCs*) [2, 3, 16, 30, 44, 62]. For example, Google halved its average energy overhead from 21% (PUE = 1.21) in 2008 to 10% (PUE = 1.10) [20]. Worldwide average PUE fell from 2.50 in 2008 to 1.57 [52] as users closed their older data centers and switched to WSCs in the cloud [34].

The relative energy consumption is then 2 × 1.57 ÷ 1.10 or 2.85x more energy (kWh) on a contemporary DSA in an average on-premise data center versus TPU v4 in Google Cloud.

4. *Map* factors in the cleanliness of the energy supply, which varies considerably by location. WSCs can be placed anywhere, while on-premise data centers depend on the local grid. For example, the average portion of *carbon free electrical energy* (*CFE*) in 2021 for the US was 40%, but it was 88% for Google's Oklahoma data centers [19]. They exploit the plentiful local renewable energy [56]. Fortunately, Oklahoma hosts all TPU v4s in Google Cloud. The global average conversion factor from electricity to *CO2 equivalent emissions—CO2e*, including greenhouse gasses like methane—is 0.475 kg/kWh [24]. After acquiring renewable energy in Oklahoma, matched on an hourly basis with our energy consumption, it dropped to 0.074.

The estimated operational CO2e for a contemporary DSA in an average on-premise data center is then 2.85 × 0.475 ÷ 0.074 or ~18.3x higher than training on TPU v4 in Google Cloud. The CFE for data centers depends on the local grid plus the availability and acquisition of renewable energy. Given such a large impact —crucial for anyone building or using ML infrastructure—it's fortunate that Google has WSCs with very high CFE to house TPU v4 supercomputers.

### 7.7 How fast do ML workloads change?

Table 1 above shows the rapid change for production workloads at Google. Note the drop in RNNs. Like RNNs, Transformers are popular for natural language translation and text summarization, but unlike RNNs they process the input all at once rather than sequentially. This revised model architecture means the operations can occur in parallel, which in turn means Transformer models can process much larger data sets. Two years after the Transformer paper was published [58], it was >20% of the TPU v3 workload. The Transformer models BERT and GPT-3 have also changed the workload landscape. Two years after the BERT paper was published [14], it was >25% of Google's workload on TPU v4, and it remained significant in 2022. Two years after publication of GPT-3 [6], LLMs were >30% of the TPU v4 production workloads. ML workloads can change dramatically in the two or more years it takes to design, build, and deploy a new ML supercomputer.

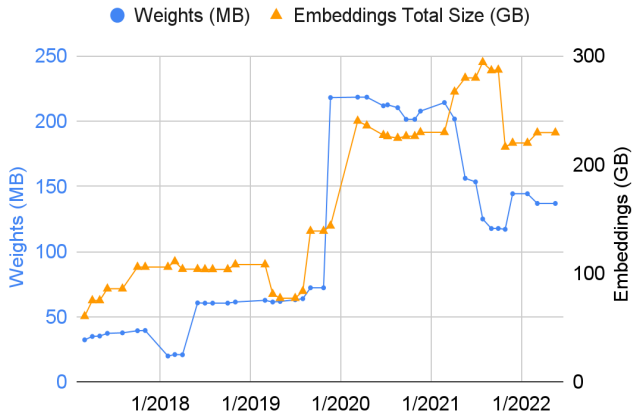

**Figure 17: Change in size of DLRM0 over time measured in weights and embeddings. Each point is a new version of DLRM0 (43 total). Each weight is 1 byte and each embedding is 4 bytes.**

### 7.8 Do individual DNN models also change?

Figure 17 shows the change in weights and embeddings for DLRM0 from 2017 to 2022. Weights grew 4.2x and embeddings grew 3.8x. Over those five years a new version was released every ~6 weeks (43 total). DLRM0 ran on all five TPU products over those five years. The velocity of change of the models *and* of the architecture highlights the importance of having a compiler that efficiently leverages the features of the underlying DSA.

### 7.9 Is MLPerf's DLRM benchmark (Figure 14 above) realistic?

Production DLRM workloads scale much better [64] than MLPerf DLRM for a few reasons. MLPerf DLRM has <2M FP32 weights while DLRM0 has 137M Int8 weights (see Figure 17). Second, the global batch size of MLPerf DLRM is capped at 64k for optimal model quality for this data set and optimizer, limiting batch size to 128 per SC on a 128-chip system (128 chips × 4 SCs/chip × 128 = 64k). In contrast, internal recommendation models often reach batch sizes of 2048–4096 and usefully scale up to 1024 chips (see Figure 11 above). Third, MLPerf DLRM has only 26 univalent features (compared to 100s in internal workloads) and no multivalent features. For these reasons, fixed overheads per batch such as HBM latency and CISC instruction generation time on the SC core sequencer are much higher on MLPerf DLRM than production workloads. Overheads like these limit its useful scalability to ≤128 chips for TPU v4 *and* A100.

### 7.10 TPU v4 has less HBM capacity than A100; could that limit LLM performance?

Our autoML LLM configuration search (Section 4) considers HBM capacity, ICI-connected aggregated FLOPS/HBM bandwidth, and other model parameters (such as batch size). The HBM capacity could be a limiting factor in some cases, but typically TPUv4 enables larger models to be partitioned across more chips with effective compute-communication overlap [59] with little overhead (for example, two case studies in Table 3 above). However, given the higher HBM capacity but smaller NVlink-connected domain of Nvidia GPUs, Nvidia's best LLM configuration might be very different from the best one for TPU v4.

### 7.11 How can DSAs avoid overspecialization?

Building a robust DSA with a multi-generational lifetime is a balancing act between domain specialization and generality with flexibility. This harmony is especially important given ML is such a fast evolving field (see Sections 7.7 and 7.8). For example, had we devoted significant resources for specialized acceleration of RNNs, that effort would have been of little use when RNN popularity plummeted. As a positive example, by providing a flexible, general, and balanced design, TPU v4 proved to be an excellent match to LLMs that were popularized [Bro22] after it was deployed.

## 8 RELATED WORK

TPU v4 uses a dedicated 3D torus interconnect. Traditional supercomputers also employ tightly connected multiprocessors over 3D tori with a high-bandwidth interconnect [15, 50]. Nvidia GPU-based systems today use a two-tier network hierarchy with NVLink and NVSwitch among groups of 4 to 256 GPUs and Infiniband beyond that.

Twisted tori are not a recent invention. The ILLIAC-IV twisted one dimension of its wrap-around links for the 2D torus that it used [5]. Sequin introduced doubly-twisted tori for mapping binary trees onto processor arrays [47]. Camara, Moreto, et al.

made the case for twisting 3D tori [7], and TPU v4 follows the k×2k×2k configuration from Camarero, Martinez, and Beivide [8].

Shalf et al. and Kamil et al. proposed a mix of circuit switching and packet switching for use in traditional supercomputers [49,28] and Kamil et al. later suggested that a MEMS-based OCS could provide circuit switching [29]. A recent paper has a similar investigation plus topology and parallelization co-optimizations to accelerate ML [60]. For data centers, there have been many proposals for OCS-based networks [17, 31, 35, 51, 53, 57]. Some even include the concept of topology engineering. However, all of this related work are paper designs or proof-of-concept, testbed scale demonstrations, in contrast to the widespread deployment of OCSes at Google for data center networks and for supercomputer interconnect. We believe that TPU v4 is the first commercial supercomputer built using OCS and the first supercomputer built with a reconfigurable interconnect that enhances performance.

The TPU v4 supercomputer implements a logically shared address space across physical chips. Software explicitly controls access and data movement; remote memories are available through asynchronous DMA writes only. The Cray T3E enabled a similar logically shared address space, with bulk asynchronous reads and writes, load-store access to remote memories, and a rich suite of atomic operations [46]. The TPU v4 memory system is tailored for high performance, with each chip maintaining tens of thousands of outstanding memory requests (to both local and remote memories).

Acceleration of embeddings is key for DLRMs used for business-critical applications, and we believe Google was the first to include on-ASIC hardware support in TPU v2, deployed in 2017. Neo from Facebook (Meta) [37] trains embedding tables with up to 12T parameters using a 128-GPU system. Neo also exploits table, row, column and data parallelism; overlaps communication and compute; and improves kernel fusion. Nvidia's MLPerf entries use similar techniques, e.g., custom fused kernels for reduction over Infiniband. Two recent papers present other optimizations for embeddings [18, 48].

## 9 SUMMARY

Two major architectural features of TPU v4 have small cost but outsized advantages. The SparseCore accelerates embeddings of DLRM models by 5x-7x by providing a dataflow sea-of-cores architecture that allows embeddings to be placed anywhere in the 128 TiB physical memory of the TPU v4 supercomputer. This gain comes at the cost of only ~5% in die area and power.

The OCSes and underlying optical components at the heart of the TPU v4 supercomputer are relatively inexpensive at <5% of overall costs and <3% of overall power consumption, yet it provides a remarkable set of eight benefits:

1. Scalability.
2. Improved availability, which enables the TPU v4 supercomputer to be 4x larger than TPU v3.
3. Modularity, allowing the faster 3D torus topology from 64 to 3072 chips and novel shapes like twisted tori.
4. Higher performance, as users can pick the topology that is best for their application.
5. Diminished power, as MEMS optical circuit switching is more energy efficient than electronic packet switching.
6. Simplified scheduling to improve utilization.
7. Faster deployment, for better return on investment.
8. Enhanced security, which encourages different organizations to share use of TPU v4 supercomputers.

Moreover, replacing OCS and ICI with Infiniband increases costs, raises power consumption, and degrades performance.

TPU v4 is faster and lower power than contemporary DSA chips made using similar technologies deployed close to the same time and for similar sized systems. The power edge might be even larger if the interconnects are included.

Training time of LLMs is greatly reduced over TPU v3 by using 3K TPU v4 slices with their 3D torus topology. The performance, scalability, and availability make TPU v4 supercomputers the workhorses of large language models (LLMs) like LaMDA, MUM, and PaLM [54, 38, 9]. These features allowed the 540B parameter PaLM model to sustain a remarkable *57.8% of the peak hardware floating point performance over 50 days* while training on TPU v4 supercomputers [9].

Google has deployed dozens of TPU v4 supercomputers, including eight for external use via Google Cloud. Moreover, the large size of the TPU v4 supercomputer and its reliance on OCSes looks prescient given that the design began *two years before* the paper was published that has stoked the enthusiasm for LLMs [6].

Advances by computer architects in the state-of-the-art of warehouse scale computing (WSCs) save energy and thus help reduce the carbon footprint of ML. When energy-efficient TPU v4 supercomputers are housed inside energy-efficient WSCs that rely on ~90% carbon free electricity, they can consume only ~⅙–½ of the energy and produce only ~5% of the operational $CO_2e$ from training on a typical contemporary ML DSA in the average on-premise data center. A ~20x reduction in carbon footprint greatly increases the chances of delivering on the amazing potential of ML in a sustainable manner [42].

## ACKNOWLEDGEMENTS

Special thanks go to the Google Platforms Optics team that kept advancing the state-of-the-art of OCSes and optical transceivers after other organizations gave up. We also thank John Hennessy, Robert Hundt, Christos Kozyrakis, Sridhar Lakshmanamurthy, Jae W. Lee, Hong Liu, Aamer Mahmood, Partha Ranganathan, Adrian Sampson, Amin Vahdat, Amir Yazdanbakhsh, George Yuan, and the anonymous ISCA reviewers for their feedback and suggestions on this paper.

We finally wish to thank our collaborators who helped with the measurements for this paper—Ben Albrecht, Jinliang Wei, Shibo Wang, Yuechao Pan, and Ziqiang Feng for Table 3 and Lluis-Miquel Munguia, Hao Wu, and Yuechao Pan for Table 6—and the many engineers who developed the TPU v4 chip, hardware, software, and the many teams contributing to its deployment, including but not limited to: Zach Cotton, Pedram Dashti, Bill Edwards, Hema Hariharan, Chetan Kale, Georgios Konstadinidis, Alan Kulawik, Justin Lee, Hong Liu, Erji Mao, Omkar Pathak, Erick Tuttle, Daoyi Wang, Kevin Yasumura, and Sara Zebian.